# Thin-Film Lithium Niobate Acoustic Resonator with High $Q$ of 237 and $k^2$ of 5.1% at 50.74 GHz

Jack Kramer[†], Vakhtang Chulukhadze[†], Kenny Huynh[‡], Omar Barrera[†], Michael Liao [‡], Sinwoo Cho[†], Lezli Matto [‡],
Mark S. Goorsky[‡] and Ruochen Lu[†]
[†]Department of Electrical and Computer Engineering, The University of Texas at Austin, Austin, US
[‡]Department of Electrical Engineering, University of California Los Angeles, Los Angeles, US
kramerj99@utexas.edu

*Summary*— This work reports a 50.74 GHz lithium niobate (LiNbO$_3$) acoustic resonator with a high quality factor ($Q$) of 237 and an electromechanical coupling ($k^2$) of 5.17% resulting in a figure of merit (FoM, $Q·k^2$) of 12.2. The LiNbO$_3$ resonator employs a novel bilayer periodically poled piezoelectric film (P3F) 128º Y-cut LiNbO$_3$ on amorphous silicon (a-Si) on sapphire stack to achieve low losses and high coupling at millimeter wave (mm-wave). The device also shows a $Q$ of 159, $k^2$ of 65.06%, and FoM of 103.4 for the 16.99 GHz tone. This result shows promising prospects of P3F LiNbO$_3$ towards mm-wave front-end filters.

## I. INTRODUCTION

Mobile communication systems rely heavily on acoustic devices for sub-6 GHz front-end filtering and signal processing [1]. This success is largely due to the intrinsic benefits that acoustic devices offer over electromagnetic alternatives, specifically significantly reduced size and loss [2], [3]. Achieving high-performance acoustic resonators at millimeter wave (mm-wave) would allow for the deployment of compact filters for emerging applications [4]. However, scaling acoustic devices to higher frequencies has posed a significant challenge, with a large performance degradation occurring above 6 GHz (survey in Fig. 1). In order to synthesize filters, higher quality factors ($Q$) and electromechanical coupling ($k^2$) are desired to increase the figure of merit (FoM, $Q·k^2$) of resonators at mm-wave. Promising material candidates for accomplishing this include thin film lithium niobate (LiNbO$_3$) [5], [6], lithium tantalate (LiTaO$_3$) [7], aluminum nitride (AlN) [8] and scandium aluminum nitride (ScAlN) [9]. While these material platforms offer intrinsically high $k^2$ and Q, achieving high-performance devices at mm-wave still poses a significant challenge.

Previously demonstrated scaling of acoustic device topologies was typically accomplished through direct reduction of the critical device dimension. In the cases of surface acoustic wave (SAW) or bulk acoustic wave (BAW) devices, this involves scaling the interdigitated transducer (IDT) pitch [10] or film thicknesses [11], respectively. This method poses significant challenges, as reduced lateral dimensions introduce fabrication difficulties as resolving ultra-fine features begins to pose a significant challenge. Similarly, ultra-thin films suffer from having a decreased volume relative to surface area scaled loss mechanisms, resulting in lower device $Q$. Thus, a device topology that allows for larger lateral feature sizes and retains thicker film thickness is highly desired.

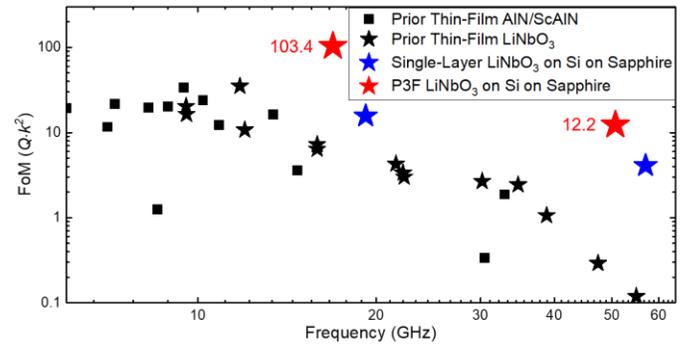

Fig. 1. Survey of state-of-the-art [18]–[20] acoustic resonator figure of merit (FoM) from 6 GHz to 60 GHz, with this work (red, bilayer P3F) and recent success (blue, single layer) using LiNbO$_3$ on Si on Sapphire highlighted.

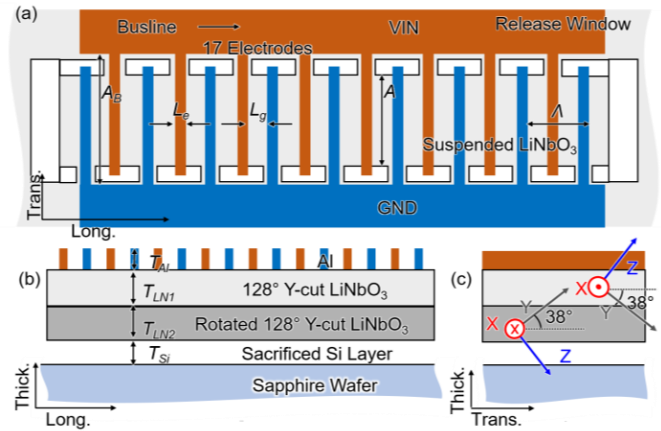

Fig. 2. (a) Top (b) cross-sectional, and (c) side schematic representations of the proposed P3F acoustic resonator design utilizing a bilayer P3F 128º Y-cut LiNbO$_3$ on a-Si on sapphire film-stack.

One method by which scaling can be accomplished is to use higher-order antisymmetric Lamb modes in suspended thin-film LiNbO$_3$ [12]–[14]. This platform offers the benefit of having a very high intrinsic $k^2$ for a 128º Y-cut LiNbO$_3$, which allows for the deployment of very wideband devices. While directly frequency scaling this device requires the thinning of the film, similar to in a film bulk acoustic resonator, one can avoid this by using higher order antisymmetric modes. As an example, instead of operating in the first-order antisymmetric (A1), half-wavelength case, one can use the third-order antisymmetric (A3) mode. However, this comes with the trade off of a reduction in $k^2$ due to a cancelation in the electric field associated with an alternating displacement pattern [15]. To help mitigate this reduction in the coupling, a periodically poled piezoelectric film (P3F) stack can be employed. This topology allows for the symmetry of the device to be changed as to support the use of higher order modes. In the case of a bilayer P3F stack, the fundamental supported mode becomes the second-order symmetric (S2) mode. Since this consists of two A1 modes supported in each layer individually, the resultant S2 mode $k^2$ is the same as the A1 mode in a single-layer film. Thus, the P3F platform enables higher order modes while maintaining high coupling, a thicker total film stack, and large lateral feature sizes. Devices based on alternating orientation layers have been previously demonstrated with strong performance [16], [17], but the benefits have yet to be demonstrated at higher frequencies.

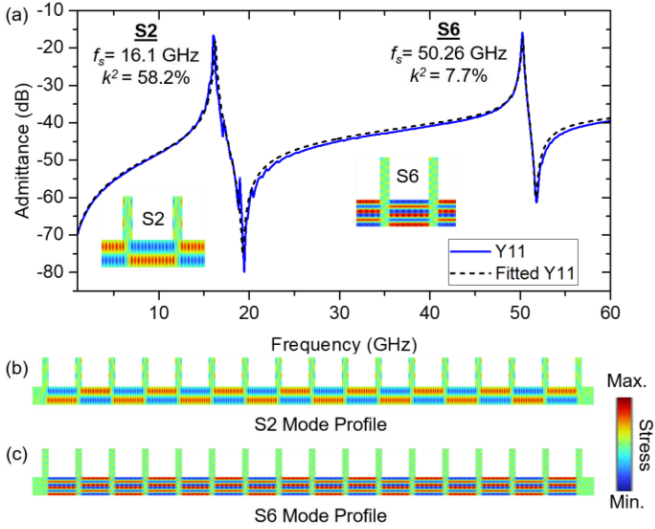

Fig. 3. (a) Simulated resonator admittance curves for the proposed film-stack with S2 and S6 tones at 16.1 GHz and 50.26 GHz respectively. (b), (c) Simulated mode stress profiles for the S2 and S6 tones, showing well confined acoustic modes within the film.

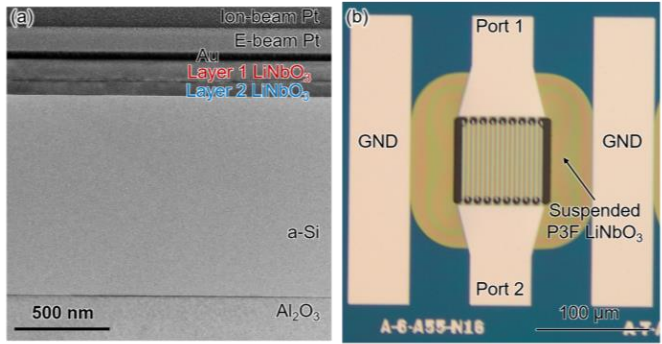

Fig. 4. (a) Transmission electron microscope (TEM) of the cross section of the film-stack. (b) Optical image of the acoustic resonator structure.

By employing the use of a novel film stack and the use of higher order second and sixth symmetric (S2 and S6, respectively) bulk acoustic tones in bilayer piezoelectric film (P3F) film stack, we are able to demonstrate record-breaking FoM performance at mm-wave.

## II. DESIGN & SIMULATION

The proposed design is shown schematically in Fig. 2 (a)-(c). The design centers around a bilayer periodically poled P3F film stack with two bonded 128º Y-cut films. The two films are rotated 180º degrees relative to each other about the axis defined by the intersection of the YZ-plane and the 128º Y-cut plane. These layers sit upon a sacrificial layer of 1 µm of amorphous silicon (a-Si) on a 500 µm sapphire wafer. This platform is an iteration on a single layer $LiNbO_3$ platform, highlighted in blue in Fig. 1, which demonstrates the promise of the substrate and intermediate layer selection. The acoustic transducer consists of an interdigitated transducer (IDT) with 17 electrodes, with an electrode width ($L_e$) of 800 nm, an electrode gap ($L_g$) of 3.2 µm, and a wavelength ($\Lambda$) of 8 µm. The aperture ($A$) of the device is 59 µm, and 350 nm thick aluminum electrodes are selected. Etch windows 4 µm in height are positioned between the electrodes with a 1 µm gap between the edge of the electrode and the etch window. Etch windows 5 µm wide define the edge of the device, and the distance between the buslines ($A_d$) of the device is 71 µm. The device is configured for a two-port measurement using 100 µm pitch GSG RF probes, with two 50 µm wide ground planes on the side of the device.

The proposed design was simulated using COMSOL finite element analysis (FEA) using an assumed $LiNbO_3$ thickness of 110 nm and Q of 200 for both layers. The admittance of the simulated device is shown in Fig. 3 (a), which showed an S2 tone at 16.1 GHz with a $k^2$ of 58.2% and an S6 tone at 50.26 GHz with a $k^2$ of 7.7%. The simulated mode profile for the S2 and S6 tones are well defined and well-constrained between the electrodes, as shown by the stress profile in Fig. 3 (b) and (c). Device parameters were optimized for the S6 tone, resulting in degraded S2 performance. However, the device still shows large $k^2$ at both tones.

## III. FABRICATION AND RESULTS

The film stack was provided by NGK Insulators Ltd. and a film stack characterization was performed on a wafer edge piece prior to fabrication. A transmission electron microscope (TEM) cross-sectional image of the stack is shown in Fig. 4 (a). The TEM shows variation in the film thicknesses between the two $LiNbO_3$ layers, which is expected to contribute to spurious mode responses away from the two primary design resonances shown in the simulation. Using the TEM, the first layer thickness is measured to be 105 nm and the second layer to be 80 nm. However, due to the thickness variation across the wafer, the actual layer thickness is expected to be slightly different for the portion of the wafer used for the fabrication.

The fabricated resonator is shown in Fig. 4 (b). First, the two layers of $LiNbO_3$ are etched via photolithographic patterning and ion milling. Then, the 350nm thick aluminum (Al) electrode and busline features are patterned using e-beam lithography and deposition. Finally, the $LiNbO_3$ is released from the substrate with a xenon difluoride ($XeF_2$) based isotropic etch. This allows for mechanical isolation of the resonator from the substrate, greatly reducing mechanical losses.

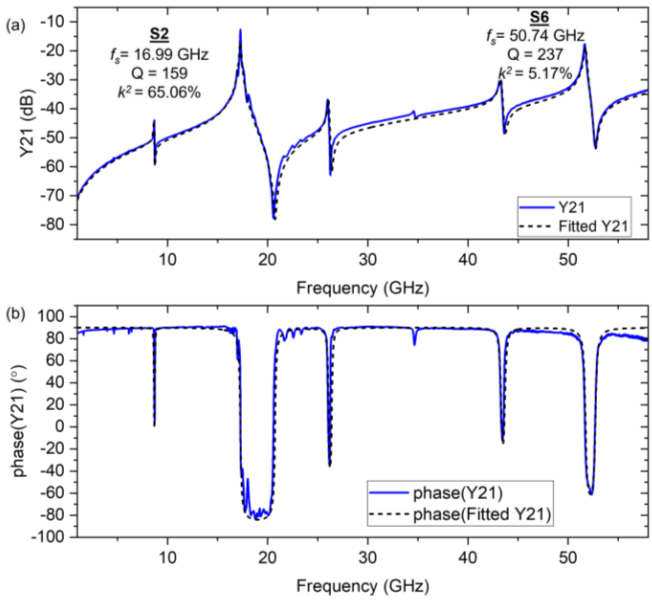

Fig. 5. (a) Measured admittance parameters and (b) phase of the fabricated device.

The measured device admittance and phase curves are shown in Fig. 5 (a) and (b). The data was fitted with the use of a five motional branch modified Butterworth Van-Dyke model, which allowed for the fitting of the two primary S2 and S6 tones as well as three main spurious modes. The measured device shows state-of-the-art performance, with a 3 dB Q of 159 and $k^2$ of 65.06% for the 16.99 GHz S2 tones and a Q of 237 and $k^2$ of 5.17% for the 50.74 GHz S6 tone. The resultant FoM for the S2 and S6 tones are 103.4 and 12.2 respectively. The data shown and performance metrics both use raw data without any de-embedding. Comparison to SoA is plotted in Fig. 1, surpassing prior works at corresponding frequencies.

IV. CONCLUSION

This work presents a SoA 50.74 GHz $LiNbO_3$ acoustic resonator with a high 3 dB Q of 237 and $k^2$ of 5.17% for the targeted S6 mode. The associated S2 mode also significantly improves over state-of-the-art works, with a Q of 159 and $k^2$ of 65.06%. The record-breaking performance is enabled using a bilayer P3F stack utilizing a sacrificial amorphous Si layer on a sapphire substrate. This work demonstrates the possibility of achieving high FoM resonators at mm-wave, enabling the future synthetization of compact mm-wave filters.

ACKNOWLEDGMENT

The authors thank the DARPA COFFEE program for funding support and Dr. Ben Griffin for helpful discussions.